\documentclass[aps,prc,reprint,onecolumn,nofootinbib,preprintnumbers,notitlepage,11pt]{revtex4-1}
\usepackage{hyperref}
\pdfoutput=1 
\usepackage{amsmath,amssymb,amsfonts}
\usepackage{graphicx}
\usepackage{booktabs,colortbl}
\usepackage{cancel,xcolor,multirow}
\usepackage[normal]{caption}
\usepackage{subcaption}
\usepackage{slashed}
\usepackage[utf8]{inputenc}
\usepackage{subcaption}
\usepackage{tablefootnote}
\usepackage{array}
\usepackage{threeparttable}

\usepackage{color}

\newcommand{\isotope}[2]{\ensuremath{^{#2}\text{#1}}}

\newcommand{\fesixty}{\ensuremath{^{60}\text{Fe}}}
\newcommand{\fe}{\ensuremath{^{56}\text{Fe}}}
\newcommand{\cosixty}{\ensuremath{^{60}\text{Co}}}

\newcommand{\nisixty}{\ensuremath{^{60}\text{Ni}}}

\usepackage{pgffor}

\newcommand{\be}{\begin{equation}} 
\newcommand{\ee}{\end{equation}} 
\newcommand{\bea}{\begin{eqnarray}}  
\newcommand{\eea}{\end{eqnarray}}
\newcommand{\bs}{\begin{split}} 
\newcommand{\es}{\end{split}}

\begin{document}


\title{Activity measurement of {\fesixty} through the decay of \isotope{Co}{60\text{m}} and confirmation of its half-life}

\author{Karen Ostdiek\footnote{E-mail: kchambe1@nd.edu}}
\author{Tyler Anderson}
\affiliation{University of Notre Dame, Notre Dame, IN 46556, USA}
\author{William Bauder}
\affiliation{University of Notre Dame, Notre Dame, IN 46556, USA}
\author{Matthew Bowers}
\author{Adam Clark}
\author{Philippe Collon}
\affiliation{University of Notre Dame, Notre Dame, IN 46556, USA}
\author{Rugard Dressler}
\affiliation{Paul Scherrer Institute Laboratory for Radiochemistry and Environmental Chemistry, 5232 Villigen, Switzerland}
\author{John Greene}
\affiliation{Argonne National Laboratory, 9700 South Cass Avenue, Lemont, IL 60439, USA}
\author{Walter Kutschera}
\affiliation{University of Vienna, Faculty of Physics, VERA Laboratory, 1090 Vienna, Austria}
\author{Wenting Lu}
\affiliation{University of Notre Dame, Notre Dame, IN 46556, USA}
\author{Austin Nelson}
\affiliation{University of Notre Dame, Notre Dame, IN 46556, USA}
\author{Michael Paul}
\affiliation{Racah Institute of Physics, Hebrew University of Jerusalem, 91904, Israel}
\author{Daniel Robertson}
\affiliation{University of Notre Dame, Notre Dame, IN 46556, USA}
\author{Dorothea Schumann}
\affiliation{Paul Scherrer Institute Laboratory for Radiochemistry and Environmental Chemistry, 5232 Villigen, Switzerland}
\author{Michael Skulski}
\affiliation{University of Notre Dame, Notre Dame, IN 46556, USA}

 
\begin{abstract}
The half-life of the neutron-rich nuclide, {\fesixty} has been in dispute in recent years. A measurement in 2009 published a value of $(2.62 \pm 0.04)\times10^{6}$ years, almost twice that of the previously accepted value from 1984 of $(1.49 \pm 0.27)\times10^{6}$ years. This longer half-life was confirmed in 2015 by a second measurement, resulting in a value of $(2.50 \pm 0.12)\times10^{6}$ years. All three half-life measurements used the grow-in of the $\gamma$-ray lines in {\nisixty} from the decay of the ground state of $^{60}\text{Co}$ (t$_{1/2}$=5.27 years) to determine the activity of a sample with a known number of {\fesixty} atoms. In contrast, the work presented here measured the {\fesixty} activity directly via the 58.6 keV $\gamma$-ray line from the short-lived isomeric state of $^{60}\text{Co}$ (t$_{1/2}$=10.5 minutes), thus being independent of any possible contamination from long-lived $^{60\text{g}}\text{Co}$. A fraction of the material from the 2015 experiment with a known number of {\fesixty} atoms was used for the activity measurement, resulting in a half-life value of $(2.72 \pm 0.16)\times10^{6}$ years, confirming again the longer half-life. In addition, {\fesixty}/{\fe} isotopic ratios of samples with two different dilutions of this material were measured with Accelerator Mass Spectrometry (AMS) to determine the number of {\fesixty} atoms. Combining this with our activity measurement resulted in a half-life value of $(2.69 \pm 0.28)\times 10^{6}$ years, again agreeing with the longer half-life.
\end{abstract}

\maketitle

\section{Introduction}
The motivation to measure the half-life of \fesixty{} stems from its natural production solely in stellar environments and from the implications of its discovery throughout the Galaxy. Neutron-rich {\fesixty} is produced in stellar environments of high neutron densities. Environments capable of having such neutron densities are in Asymptotic Giant Branch (AGB) stars and massive stars, through the neutron-producing reactions of \isotope{C}{13}($\alpha$, n)\isotope{O}{16} and \isotope{Ne}{22}($\alpha$, n)\isotope{Mg}{25} respectively. The material produced in these environments can then be released to the surrounding interstellar medium through supernova explosions and hypothesized processes such as stellar winds. Therefore, {\fesixty} can be expelled and observed in the Galaxy specifically in three distinct ways.

\textit{$\gamma$-ray observations}: The decay of {\fesixty}, specifically two $\gamma$-rays from the decay of its daughter product, \isotope{Co}{60\text{g}}, at energies of 1173 keV and 1332 keV (see Figure \ref{fig:MyDecay1} for the full decay scheme of {\fesixty}), has been observed by 19 BGO-shielded HPGe detectors on the spacecraft INTEGRAL when looking toward the center of our Galaxy \cite{Wang2007}. This observation suggests that nucleosynthesis is an ongoing process, as {\fesixty}'s half-life is significantly shorter than the age of the Galaxy. 

$^{\textit{60}}$\textit{Ni} \textit{in meteorites}: Lower {\nisixty}/\isotope{Ni}{58} isotopic ratios, {\nisixty} being the granddaughter of {\fesixty}, have been found in meteorites as compared to samples from Earth and Mars \cite{Bizzarro2007}. These meteorites would have been formed during the early Solar System (ESS). The higher isotopic ratios in younger samples supports the hypothesis that {\fesixty} was injected into the Solar System after its formation. Precise timing and abundances of {\fesixty} in our Solar System would put constraints on ESS models and the environment in which it formed.

$^{\textit{60}}$\textit{Fe} \textit{excesses in ocean crust, lunar, and microfossil samples}: Studies on ocean crust samples have found an excess of \fesixty{} above background levels, dating to approximately 1.5-3.2 million years ago (\cite{Knie2004}, \cite{Fitoussi2008}, \cite{Wallner2016}) as well as 6.5-8.7 million years ago \cite{Wallner2016}. Similar signatures have been found in lunar samples \cite{Fimiani2014} and in microfossil records \cite{Bishop2012}. These excesses would seem to indicate that the Solar System passed through the debris field of multiple supernova events in the last 10 million years, as discussed in \cite{Breitschwerdt2016}. 

The first half-life value, published in 1957 by Roy and Kohman \cite{RoyKohman1957}, was $3\times10^{5}$ years with a factor of 3 uncertainty. After this publication, it was determined that certain assumptions made in it, specifically the relative production rates of {\fesixty} versus \isotope{Fe}{59}, may have been incorrect. Therefore a longer half-life value could not be ruled out. Kutschera et al. measured the half-life in 1984, finding $(1.49\pm 0.27)\times10^{6}$ years by a combination of an activity and an Accelerator Mass Spectrometry (AMS) measurements \cite{Kutschera1984}. However, the complex AMS experiment may have resulted in a somewhat lower {\fesixty} isotopic ratio in the sample material than was actually present. In a third measurement by Rugel et al. in 2009, the {\fesixty} isotopic ratio was determined from new sample material with a higher {\fesixty} isotopic ratio by multicollector-inductively coupled plasma mass spectrometry (MC-ICPMS). This measurement published a significantly longer half-life of $(2.62\pm0.04)\times10^{6}$ years \cite{Rugel2009} and was confirmed in 2015 by Wallner et al., which published a value of $(2.50 \pm 0.12) \times 10^{6}$ years \cite{Wallner2015}.

The last 3 half-life measurements have all used the decay of the ground state of {\cosixty}  to quantify the activity of an {\fesixty} sample, coupled with a measurement of the number of {\fesixty} atoms. This work, in contrast, focuses on the use of the direct decay of the isomeric state of {\cosixty}, as did the Roy and Kohman measurement \cite{RoyKohman1957}. The sample used for this work is described in the following section. In Sections \ref{Activity} and \ref{AMS}, the experimental procedures are discussed, including the decay scheme of {\fesixty}, the direct decay activity measurement, and a determination of the number of {\fesixty} atoms in the sample using Accelerator Mass Spectrometry.  

Uncertainties in this work were estimated according to the recommendations in the Guide to Expression of Uncertainty in Measurements \cite{GUM2008}. All given uncertainties are combined standard uncertainties with a coverage factor of k=1.

\section{Sample Material}\label{Sample}

\fesixty{} is only naturally produced in slow neutron capture process sites such as stellar environments. The samples used for this work, produced artificially, come from spallation reactions resulting from high energy (590 MeV) protons incident on a copper beam stop at the Paul Scherrer Institute. The beam stop was in use for 12 years, building up numerous radioactive isotopes including \fesixty{} \cite{Schumann2013}. Material was extracted from the beam stop and iron was chemically separated. The material of this present work was originally used as a target in a cross section measurement of \fesixty{}(n,$\gamma$)\isotope{Fe}{61} at stellar energies \cite{Uberseder2009}. The {\fesixty} from the target was later recovered and {\cosixty} was chemically removed. Some of this recovered material was sent to the Vienna Environmental Research Accelerator (VERA) Laboratory in Austria to be used to create a dilution series. In this series, a total of four samples were created, each with an {\fesixty}/{\fe} isotopic ratio subsequently lower by one order of magnitude.  Further details  of each sample can be found in Wallner et al. \cite{Wallner2015}.  By knowing the amount of stable iron added to the sample and using the technique of AMS on a small subset of the sample to determine its isotopic ratio, the total number of {\fesixty} atoms in the sample can be calculated. Portions of each of the four resulting samples from the dilution series were combusted into iron oxide powder, with the rest remaining as a liquid. 

The University of Notre Dame measurement concentrated on two of the samples, Fe-1 and Fe-4, and their expected isotopic ratios can be found in Table \ref{Samples}. For this work we received powdered versions of each sample. The powdered material of Fe-4, used for an AMS measurement, was concurrently measured by Wallner et al. \cite{Wallner2015}. We also recieved the remaining liquid part of the most concentrated sample, Fe-1. The liquid solution of Fe-1, with an identical {\fesixty}/{\fe} isotopic ratio as the powdered Fe-1 material, was evaporated into a point source for the activity measurement. Performing the AMS and activity measurements on the same material in principle bypasses the need to rely on the dilution factor, shown in Table \ref{Samples}.

\begin{table*}[h]
\centering
\caption{Dilution series created at the Vienna Environmental Research Accelerator Laboratory including the amount of added {\fe} for each sample.}
\label{Samples}
\renewcommand{\arraystretch}{1.4}
\scalebox{1.1}{
\begin{tabular}{|c|ccccc|}
\hline
\begin{tabular}[c]{@{}c@{}}Sample\\ Name \footnote{Partial table reproduction from Wallner et al. \cite{Wallner2015}.}\end{tabular} & \begin{tabular}[c]{@{}c@{}}Fe\\ Carrier\footnote{Fe standard solution with 1 mg Fe/mL.}\\ (mg)\end{tabular} & \begin{tabular}[c]{@{}c@{}}$\text{N}_{56}$\\ ({\fe} at)\\ ($\times 10^{20}$)\end{tabular} & \begin{tabular}[c]{@{}c@{}}Dilution \\ factor\\ for $^{55,60}\text{Fe}$\end{tabular} & \begin{tabular}[c]{@{}c@{}}$\text{N}_{60}/\text{N}_{56}$\\  relative \\ to Fe-1\end{tabular} & \begin{tabular}[c]{@{}c@{}}Nominal\\ Isotopic Ratio\\ ({\fesixty}/{\fe})\end{tabular} \\ \hline
Fe-1                                                  & 50.0                                                        & 4.95                                                                                      & 1                                                                                    & 1                                                                                            & $\sim 2\times10^{-6}$                                                                    \\

Fe-4                                                  & 55.55                                                       & 5.50                                                                                      & 1000                                                                                 & 0.00090                                                                                      & $\sim 2\times10^{-9}$                                                                    \\ \hline
\end{tabular}
}
\end{table*}

\section{Experimental Procedure: Activity}\label{Activity}
\begin{figure*}[t]
\begin{center}
\includegraphics[width=.75\linewidth]{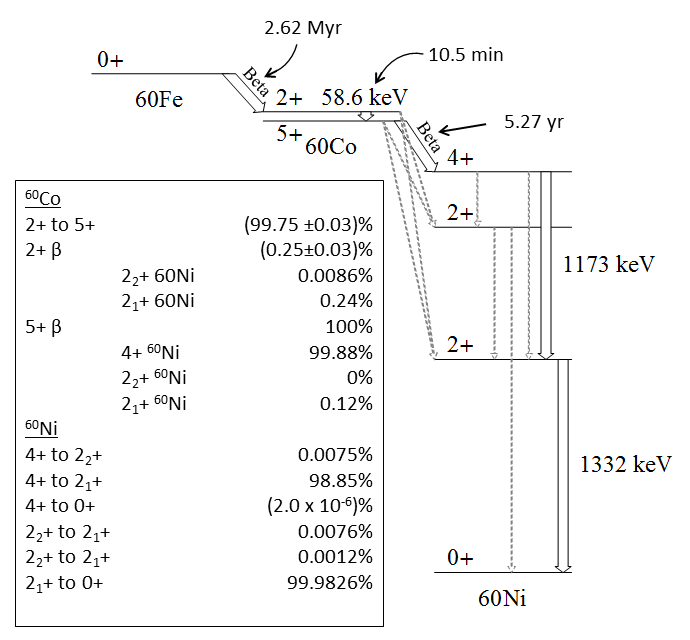}
\caption{Full decay scheme for {\fesixty}. Thick white arrows indicate decays that happen more prevalently (100\% or almost 100\% for each) and gray, dashed arrows indicate other possible decays that occur (data take from \cite{Browne2013}).}
\label{fig:MyDecay1}
\end{center}
\end{figure*}

As shown in Figure \ref{fig:MyDecay1}, \fesixty{} decays to the 2+ isomeric state in \cosixty{}. This state then primarily decays to the ground state in \cosixty{}, (99.75 $\pm$ 0.03)\% of the total, with a half-life of 10.467 minutes. The decay is either via internal conversion, ($97.93 \pm 0.03$)\%, or the emission of a $(58.603 \pm 0.007)$ keV $\gamma$-ray, (intensity, I$_{\gamma}$\% = $2.07 \pm 0.03$)\%. From here, the ground state of \cosixty{} decays to an excited state in \nisixty{} with a half-life of (1925.28 $\pm$ 0.14) days. These excited states decay quickly to the stable ground state of \nisixty{}. The predominant lines here are the cascades of $(1173.228 \pm 0.003)$ keV and $(1332.492 \pm 0.004)$  keV. Further details of the decay scheme of {\fesixty} can be found in \cite{Browne2013}. 

These last two $\gamma$-ray lines, as a proxy for the decay of {\fesixty}, are referred to as the grow-in decay because of the necessary wait time for the decay of \cosixty{}'s ground state. Measuring the grow-in decay requires and assumes a reduction of any possible {\cosixty} present in the {\fesixty} material by chemical separation to negligible levels prior to starting the activity experiment. All previous measurements of the half-life, not including the initial one in 1957, have used the grow-in decay. Conversely, this work measured the isomeric decay of \cosixty{}'s 2+ excited state, specifically the 58.6 keV $\gamma$-ray line, which eliminates the need for the wait time and complex chemistry techniques as it is directly fed by the beta decay of {\fesixty}. 

As discussed above in Section \ref{Sample}, the sample used for both the activity and the AMS (see below, Section \ref{AMS}) measurements was the Fe-1 sample. Specifically for the activity measurement, the remaining 13.0016 mL of Fe-1 was used. This sample was reduced to a point source at the Physics Division of Argonne National Laboratory by evaporating most of the HCl acid solution in the sample. Once the sample size was less than 0.1 mL, it was transferred by a loss-less pipet to a piece of Mylar (1.25 inches by 1.25 inches, 0.002 inches thick)  and allowed to dry fully forming a deposition spot of 0.242 inches in diameter. All vessels, vials, pipets, etc. were measured for activity after the evaporation. There was negligible activity left on any of the materials and all vials, which were weighed before and after the evaporation, had no changes outside of the uncertainty of the scale used.

The total {\fesixty} activity in the sample is relatively small and was expected to be about 1.3 Bq from the data given in \cite{Wallner2015}. Therefore, a close-up counting geometry using two planar, high-purity germanium (HPGe) detectors in a head-to-head arrangement was used. Planar HPGe detectors exhibit high efficiencies (on the order of 10\% for full energy peak efficiencies) at low energies (3 to 300 keV) due to the use of a thin beryllium window mounted on the end cap. 

\begin{figure}[h]
\begin{center}
\includegraphics[width=.5\paperwidth]{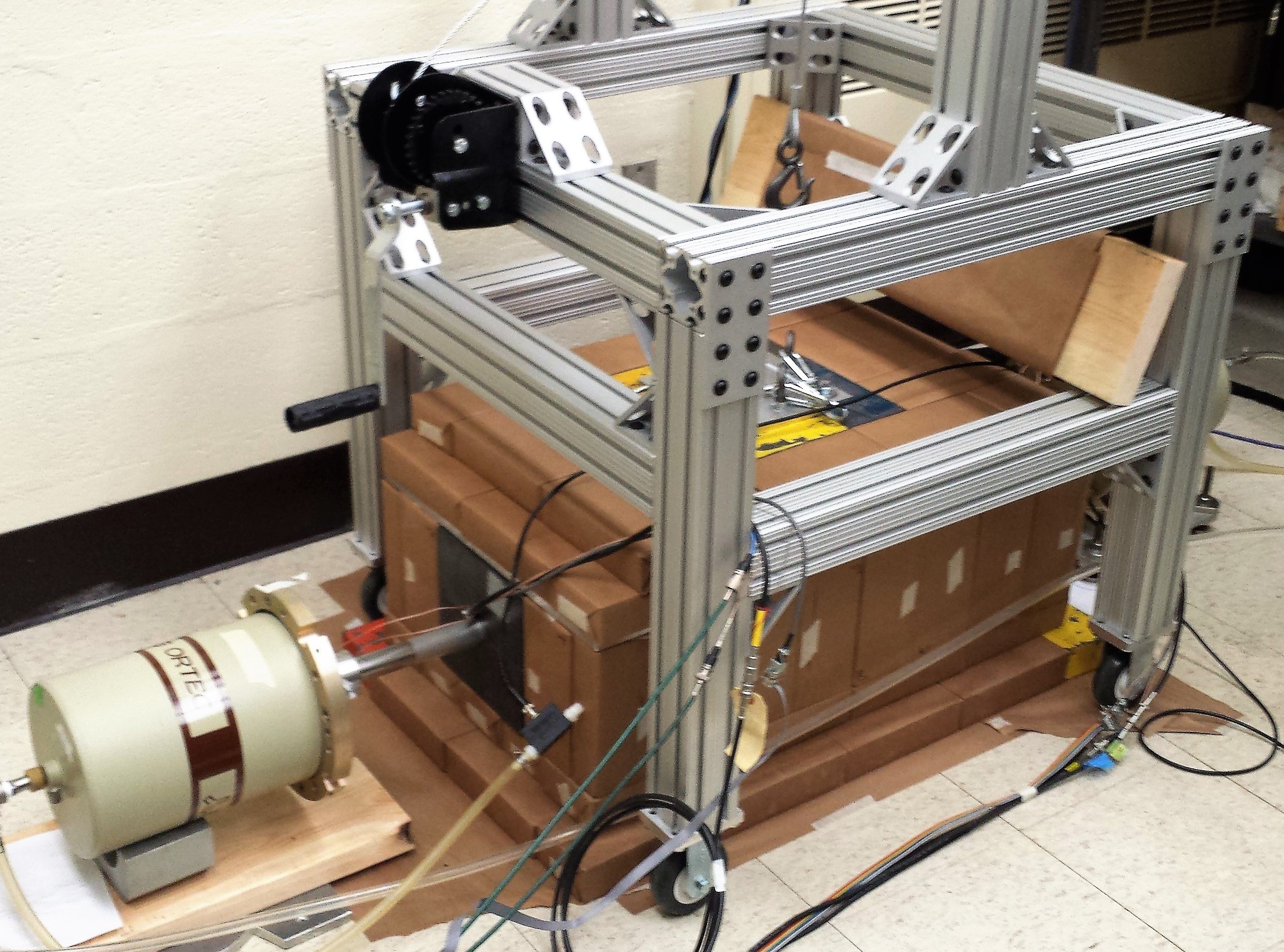}
\caption{Lead castle for the low-background counting station. Two planar HPGe detector heads fit inside of this lead castle construction, one on the left side of this picture and the other exactly opposite (not visible here). The lead castle is two layers of lead bricks thick on all sides. The aluminum structure surrounding the lead is a winch system for removing a section of the top two layers, allowing easy access to the inside and eliminating line-of-sight issues. For this work, the detector heads were (12.5 $\pm$ 0.25) mm from the target (or (17.5 $\pm$ 0.25) mm for the last measurement) with a plastic target holder centered between them, which would hold samples and calibration sources at the same location.}
\label{fig:Castle}
\end{center}
\end{figure}

Both detectors were ORTEC Model GLP 50XXX/15-S, with the following characteristics: Active crystal diameter: 51.0 mm, Active crystal depth: 14.3 mm, Be window thickness: 0.5 mm, Crystal position from inside of Be window: 11 mm (Detector 1), 12 mm (Detector 2). The detectors were placed in a head-to-head geometry and were completely surrounded by two layers of lead bricks (for a total 10 cm wall thickness), specifically selected for their low intrinsic background activity. This was done to suppress the environmental background. A photograph of the lead castle configuration is shown in Figure \ref{fig:Castle}. The advantage of this setup is the enhanced registered count rate of the {\cosixty} $\gamma$-rays.

For the efficiency calibration, a 1 mL aliquote of a certified \isotope{Am}{241} solution from Eckert and Ziegler, (3763 $\pm$ 113) Bq, was used. This particular isotope was chosen because of a predominant $\gamma$-ray line at 59.54 keV (intensity of (35.9 $\pm$ 0.07)\%) which is within the full-width at half-maximum of the detectors to the emission line of \isotope{Co}{60\text{m}}

To cancel out additional corrections due to differences in the Fe-1 sample's geometry, attenuations factors of the backing material, and the chemical composition, the \isotope{Am}{241} reference source needs to be of a similar geometry and composition. As the Fe-1 sample has 6.5 mg of stable iron in it (13\% of the 50 mg added to the total Fe-1 sample), the same amount was added to the \isotope{Am}{241} prior to evaporation (see Table \ref{tab:Amounts}). The \isotope{Am}{241} source then went through the same evaporation process as the Fe-1 activity sample so that both would have very similar properties.

\begin{table*}[h]
\centering
\caption{Information on the Fe-1 sample and the \isotope{Am}{241} source in the initial conditions.}
\label{tab:Amounts}
{\renewcommand{\arraystretch}{1.5}%
\begin{tabular}{|l|c|c|}
\hline
Sample & Isotope        &  Amount of Atoms of Interest          \\ \hline
Fe-1 (total)                   &{\fesixty} &  $1.145 \times 10^{15}$ \footnote{Amount of {\fesixty} determined by Wallner et al. \cite{Wallner2015}}  \\
Fe-1 (13\% of total) &{\fesixty} & $1.495 \times 10^{14}$ \footnote{Calculated as 13.0016\% of the amount in Fe-1 (total).} \\
\isotope{Am}{241} standard         & \isotope{Am}{241}&  $7.283 \times 10^{13}$ \footnote{Value calculated from the activity quoted by the manufacturer.} \\ \hline
\end{tabular}}
\end{table*}

Evaporation losses during the preparation process amount to less than 0.1\%. Also both the unknown Fe-1 activity sample and the \isotope{Am}{241} source were placed in identical target holders so that the distance from either detector to the sample was (12.5 $\pm$ 0.25) mm. The detection efficiency as a function of position from the detector head yields an additional variation of $\pm$2\% due to the target holder position.

\begin{figure}
\centering
\includegraphics[width=.75\textwidth]{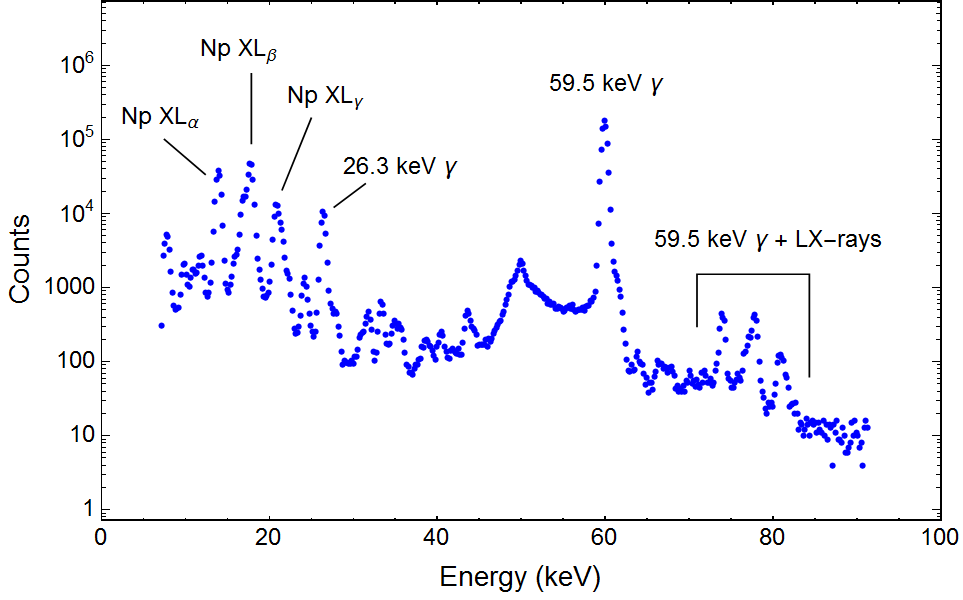}
\caption{Typical spectrum of a calibration measurement using the customized \isotope{Am}{241} reference source, on Detector 1, for 3600 seconds of live time. Counts per bin are plotted as a function of energy (keV). Several peaks of interest are labeled, in particular the \isotope{Np}{237} L X-ray peaks, the $\gamma$-rays of \isotope{Am}{241}, and the true coincidence summing peaks of the predominant $\gamma$-ray with the \isotope{Np}{237} L X-rays.}
\label{fig:CalText} 
\end{figure}

Figure \ref{fig:CalText} shows a typical calibration spectrum from 10 keV to 90 keV. The three predominant \isotope{Np}{237} L X-rays (Np XL$_{\alpha}$ $\sim$ 15.9 keV, Np XL$_{\beta}$ $\sim$ 17.8 keV, and Np XL$_{\gamma}$ $\sim$ 20.9 keV) are visible in front of the two characteristic $\gamma$-ray lines of \isotope{Am}{241} at 26.3 keV (2.27$\pm$0.12)\% and 59.54 keV (35.90 $\pm$ 0.07)\% \cite{Basunia2006}. In addition, in the range between 70 keV and 80 keV, full energy true coincidence summing peaks of the three X-rays and the main $\gamma$-ray line of \isotope{Am}{241} are clearly visible. Unfortunately, such summing effects do not only appear with full energy $\gamma$-rays but also with Compton scattered ones, making the determination of the detection efficiency more challenging. Therefore, as these true coincidence summing effects reduce the count rate of the calibration source peak, a more sophisticated calibration procedure is needed.


Both the full energy peak and the total efficiency values needed to precisely estimate the true coincidence summing correction factors for various sample-to-detector distances were calculated using the general purpose Monte-Carlo code MCNP6 v1.0 \cite{MCNP6} being interoperable with the ENDF/B-VII.1 cross-section database \cite{ENDFB71}. In general, this Monte-Carlo method is the preferred way for a numerical solution of the radiation transport equation, especially in complex geometries.

The MCNP model consists of the sample, the Mylar backing, the detector heads with the thin beryllium windows, and the HPGe crystals inside. Back-scattering of photons was only considered for the opposite detector assembly but not for the surrounding lead castle. Two independent sets of MCNP parameter studies were performed,  facing the sample deposition to Detector 1 and the Mylar backing to Detector 2 and vice versa. In each parameter study the distance of the sample varied from 0.5 mm to 24.5 mm with respect to Detector 1. Both the total and full energy peak efficiencies of the two dominant  \isotope{Am}{241}  $\gamma$-lines and all of \isotope{Np}{237} L X-rays were calculated. The statistical uncertainties of the results from individual Monte Carlo runs were smaller than 0.15\%.

 A parabolic regression analysis was then performed between the MCNP result of the TCS correction and the peak efficiency $\epsilon_{\text{Am}}\%$. The final uncertainty takes into account the fit of the parameters as well as a general uncertainty of 10\% typical for MCNP calculations applied to estimate efficiencies of HPGe detectors (see e.g. \cite{Ewa2001103}). These parabolic correlation functions were used to estimate the TCS correction based on the measured \isotope{Am}{241} peak efficiency $\epsilon_{\text{Am}}\%$, thus circumventing the problem of the exact position determination of respective measurements. An additional random uncertainty of 2\% was applied to account for possible deviations of the sample positioning when exchanging the \isotope{Am}{241} reference source with the \isotope{Fe}{60} sample. Experimental determination of the summing effects of the three X-rays with the main $\gamma$-ray (encompassing the region between 65 and 85 keV, see Figure \ref{fig:CalText}) was ($1.0 \pm 0.5$\%), accounting for about 1/5 of the total modeled correction of $\sim$ 5\%.

The efficiency Eff.\% to be applied to determine the \isotope{Fe}{60} activity is calculated from the efficiency $\epsilon_{\text{Am}}$\% of the \isotope{Am}{241} 59.54 keV line, corrected for its true coincidence summing (TCS) effect using the following equation:

\begin{center}
\begin{eqnarray}
\text{Eff.\%} = \frac{ \epsilon_{\text{Am}}\% }{\text{TCS}} 
= \frac{ \epsilon_{\text{Am}}\% }{ \left( 1 - \sum_i \nu_{x i} \cdot \tau_{x i} \right)}
\label{eq:Eff}
\end{eqnarray}
\end{center}

In Equation \ref{eq:Eff}, $\nu_{x i}$ and $\tau_{x i}$ denote the X-ray intensity and corresponding total efficiency, respectively. In contrast to \isotope{Am}{241}, the isomeric transition of \isotope{Co}{60\text{m}} is not accompanied by X-ray emissions. Therefore, the Eff.\% efficiency is used to obtain the {\fesixty} activity of the sample.

\begin{figure*}[h]
\centering
\includegraphics[width=.75\textwidth]{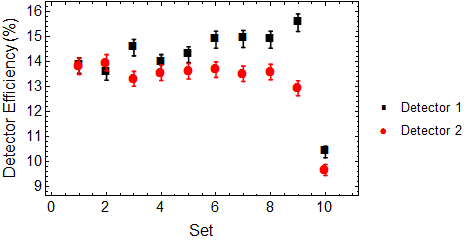}
\caption{Detection efficiencies (labeled as Eff\% in Equation \ref{eq:Eff}) for the two planar HPGe detectors used, as measured with a \isotope{Am}{241} calibration source using the 59.54 keV $\gamma$-ray and corrected for true coincidence summing effects. The percent error is 2.27\%, as detailed in Table \ref{tab:ActivityError}. Detector 1 is shown as black squares and Detector 2 is shown as red circles. The subtantial change in efficiency between sets 1-9 and set 10 comes from changing the distance between the detectors and the source. The detectors for sets 1-9 are (12.5 $\pm$ 0.25) mm from the target and for the final set, set 10, are (17.5 $\pm$ 0.25) mm from the target. For each set, the target holder was in the same location for the Fe-1 activity sample and the \isotope{Am}{241} source. For the final calculations on the Fe-1 sample, the efficiencies measured before the sample were used to scale the activity.}
\label{fig:DetectorEff}
\end{figure*}

\begin{figure}[h]
\begin{center}
\includegraphics[width=.75\textwidth]{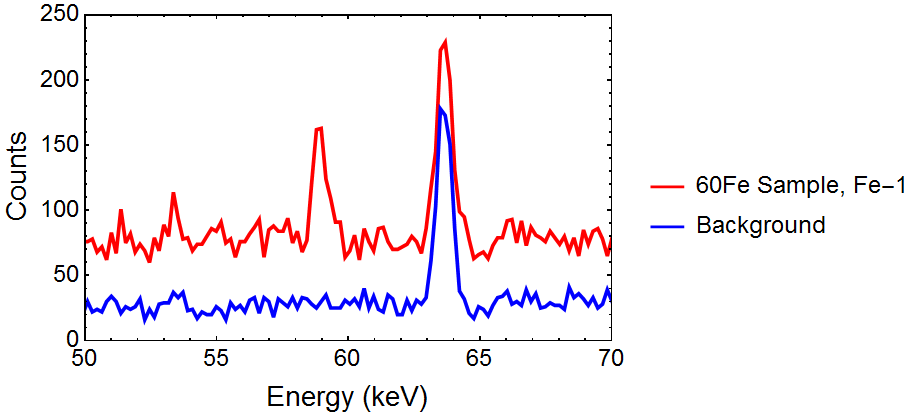}
\caption{24-hour runs on the Fe-1 sample (in red, upper line) and the background (in blue, lower line). Counts per energy bin are on the y-axis and energy in keV is on the x-axis. Note the background peak at 63.3 keV. This peak is from the decay of \isotope{Th}{234} in the \isotope{U}{238} decay chain. This peak is well separated from our peak of interest at 58.6 keV but can act as a good test of the background subtraction technique. The shift in the background continuum when the Fe-1 sample is in place is due to the internal bremsstrahlung photons from the electron capture decay of \isotope{Fe}{55} which is present in the original {\fesixty} material \cite{Wallner2015}.}
\label{fig:example}
\end{center}
\end{figure}

Efficiency measurements were performed before and after each sample run for a total of 3600 seconds of live time each. This allowed us to track any significant changes in the detectors. For the final data evaluation, the efficiencies measured before each {\fesixty} sample measurement were used and are shown in Figure \ref{fig:DetectorEff}. A sample run was conducted for 6 days, real time. Figure \ref{fig:example} shows one data set with 24 hour runs on the background and the sample.

There are three things to note about this spectrum. First, there is a background peak at approximately 63.3 keV which is predominant in both the background and the sample runs. This line comes from the decay chain of \isotope{U}{238}, present in all modern lead bricks. Second, this background peak is well separated from the energy region of interest at 58.6 keV. Thirdly, there is a significant shift in the continuum background when the Fe-1 activity sample is measured. This is due to the internal bremsstrahlung from the electron capture of \isotope{Fe}{55} (Q$_{\text{EC}}$=231.21 keV, \cite{J2008}), which is the main activity in the original {\fesixty} sample material. This continuum shift is accounted for in the background subtraction and the 63.3-keV peak acts as a check of that process. With and without the sample, the count rate of the 63.3 keV peak after background subtraction is within uncertainty.

\begin{figure}[h]
\centering
\includegraphics[width=.75\textwidth]{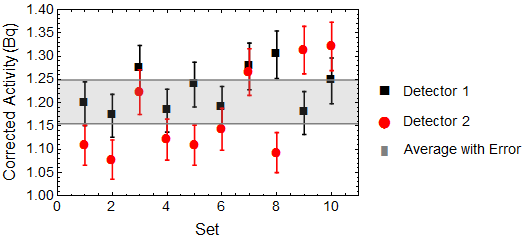}
\caption{The results of the Fe-1 activity sample where each set occurs over a 6-day real-time period. Both detectors are given here, one in black squares and the other in red circles. In the gray band, the final average number of 1.202 Bq and a 3.91\% uncertainty is shown,as calculated from Table \ref{tab:ActivityError}.}
\label{fig:ActivitySets}
\end{figure}

The activity is calculated using the following equation where Br. Ratio is the branching ratio of the 2+ excited state in \cosixty{} to the 5+ ground state in {\cosixty}, and I$_{\gamma}$ \% is the intensity of the \isotope{Co}{60\text{m}} $\gamma$ decay.
\begin{center}
\begin{eqnarray}
\text{Activity}=\frac{\text{Counts}}{\text{second}}\times \frac{100}{\text{Eff.\%}}\times \frac{100}{\text{Br. Ratio\%}}\times \frac{100}{\text{I$_{\gamma}$\%}}
\end{eqnarray}
\end{center}
The average activity, taken from the data sets of both detectors with a total live run time of more than 118 days cumulatively, of the \isotope{Co}{60\text{m}} peak is $(1.202 \pm 0.047)$ Bq. Remembering that this sample is (13.0016 $\pm$ 0.0001) g of the original 100 g sample, the activity of the original Fe-1 sample is then $(9.245 \pm 0.361)$ Bq. The systematic uncertainty budget for the activity measurement is given in Table \ref{tab:ActivityError}. The individual sets of 6-day runs on the Fe-1 activity sample and the final combined uncertainty of the activity (grey band) are shown in Figure \ref{fig:ActivitySets}.  

\begin{table}[h]
\centering
\caption{Activity Measurement Uncertainty Calculation.}
\label{tab:ActivityError}
{\renewcommand{\arraystretch}{2}
\begin{tabular}{|c|c|c|}
\hline
                                       \multicolumn{1}{|c|}{Quantity \footnote{Bold quantity is the final total.}}                         & Value $\pm$ Uncertainty                 & Percent Error (\%) \footnote{Other contributors to the uncertainty, such as statistics, are negligible comparatively.}    \\ \hline

                                      \isotope{Am}{241} $\gamma$ yield 559.54 keV                                      & (35.90 $\pm$ 0.07)\%               & 0.20                 \\
                                      Activity of the \isotope{Am}{241} source &  (3763.0 $\pm$ 37.6) Bq/mL & 1.0 \\
                                      Transfer losses of  \isotope{Am}{241}  &  $<$0.1\% & 0.1 \\  
                                      \isotope{Am}{241} Peak Area Determination & 642276 counts \footnote{Explicit vaues for measurements of Set 1 using Detector 1. \label{fn:Set1}} & 2.0 \\ 
                                       TCS Correction Factor   & (0.945 $\pm$ 0.003)\textsuperscript{\ref{fn:Set1}} & 0.29 \\ \hline \hline  
                                      \textbf{Efficiency, Eff. \%}                                   & \textbf{(13.84 $\pm$ 0.31)\%} \textsuperscript{\ref{fn:Set1}}                    & \textbf{2.27}                  \\ \hline \hline

                                       Branching Ratio of 2+ to 5+ states in {\cosixty}                                      & (99.75 $\pm$ 0.03)\%               & 0.030                 \\
                                        \isotope{Co}{60\text{m}} $\gamma$ intensity at 58.6 keV                                   & (2.07 $\pm$ 0.03)\%                & 1.45                  \\ 
                                     Aliquot of the original Fe-1 Sample                  & (13.0016 $\pm$ 0.0001) mL      & $7.7 \times 10^{-4}$ \\ 
                                      Transfer losses of  \isotope{Fe}{60}  & $<$0.1\% & 0.1 \\ 
                                       \isotope{Fe}{60} Sample Position 					& (12.5 $\pm$ 0.25) mm \footnote{The nominal value for Sets 1-9. For Set 10, the position is (17.5 $\pm$ 0.25) mm.} 		      &2.0                   \\ 
                                       \isotope{Fe}{60} Peak Area Determination & 7103 counts \textsuperscript{\ref{fn:Set1}} & 2.0 \\ \hline \hline 
                                       \textbf{Activity of Fe-1 Sample}  & \textbf{(1.202 $\pm$ 0.047) Bq}              & \textbf{3.91}                   \\ \hline \hline
\end{tabular}}
\end{table}

\section{Accelerator Mass Spectrometry and Half-life Comparison}\label{AMS}
As we had also received small amounts of all four samples in the dilution series in powder form as described in Section \ref{Samples}, it was compelling to perform a confirmation of the isotopic ratios published by Wallner et al. \cite{Wallner2015} using Accelerator Mass Spectrometry (AMS). Additionally we were able to directly measure the isotopic ratio of the Fe-1 material and did not rely purely on the factors given from the dilution series. Therefore this work is the first coupled measurement on the same material. Specifically, the material used for the activity and the AMS experiments had undergone all of the same chemistry steps, including {\cosixty} reduction. The only difference was that the AMS material was ignited into a powder whereas the activity material was evaporated.

The success of an AMS measurement hinges on the separation of a rare isotope (usually a long-lived radioisotope) from its abundant stable isotopes and isobaric interferences. {\fesixty} has both stable iron isotopes and interference from an intense stable isobar, {\nisixty}. Various techniques are employed in order to remove both of these from the main beam before particle identification. For this work, the FN accelerator at the University of Notre Dame's Nuclear Science Laboratory was used, operating at 8.5 MV and with 2 sets of carbon stripper foils. In this configuration, we produced a beam with an energy of 112.93 MeV. Isotopic and charge selections occur in the pre-acceleration 60$^{\circ}$ magnet, the post-acceleration 90$^{\circ}$ magnet, and in the Wien Filter on the AMS beam line. Spatial separation of the isobar {\nisixty} is performed using a 90$^{\circ}$ Spectrograph magnet in Gas-Filled Mode and detected with a Parallel Grid Avalanche Counter detector. Further separation uses energetic differences between the isobars through Bragg curve spectroscopy in an ionization chamber immediately following the Spectrograph magnet. Details of the facilities, detectors, and techniques used can be found in Ostdiek, et al. \cite{Ostdiek2015}.     

\begin{figure*}[h]
        \centering
        \begin{subfigure}[b]{0.475\textwidth}
            \centering
            \includegraphics[width=\textwidth]{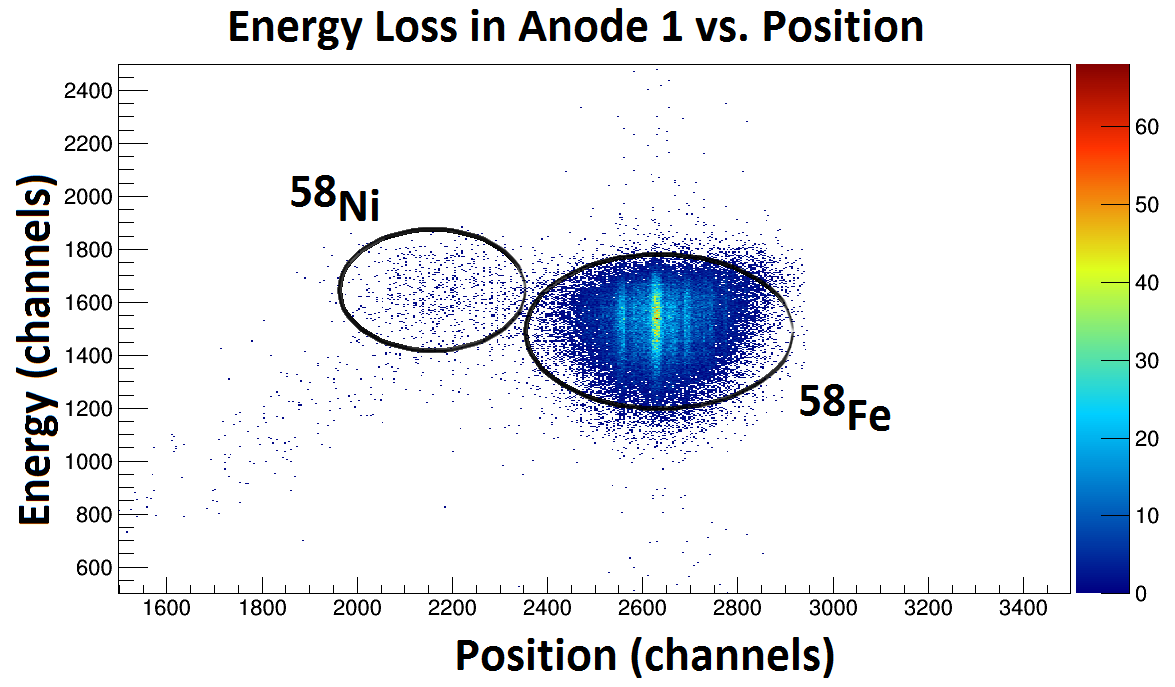}
            \caption{\small Blank dE1 vs. Position.}    
            \label{fig:58de1}
        \end{subfigure}
        \hfill
        \begin{subfigure}[b]{0.475\textwidth}  
            \centering 
            \includegraphics[width=\textwidth]{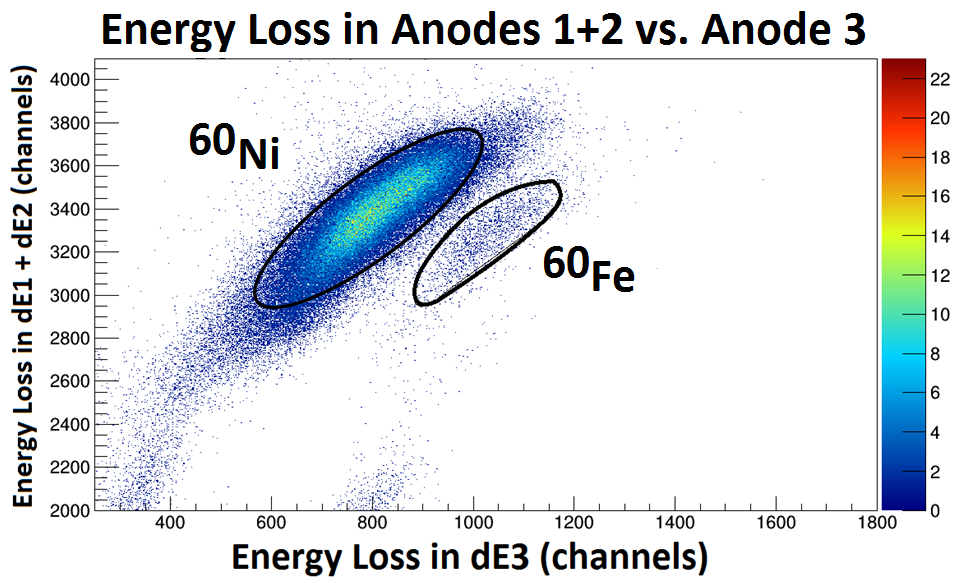}
            \caption{\small Fe-4 dE1 + dE2 vs. dE3, cuts drawn.}    
            \label{fig:Fe4drawn}
        \end{subfigure}
        \vskip\baselineskip
        \begin{subfigure}[b]{0.475\textwidth}   
            \centering 
            \includegraphics[width=\textwidth]{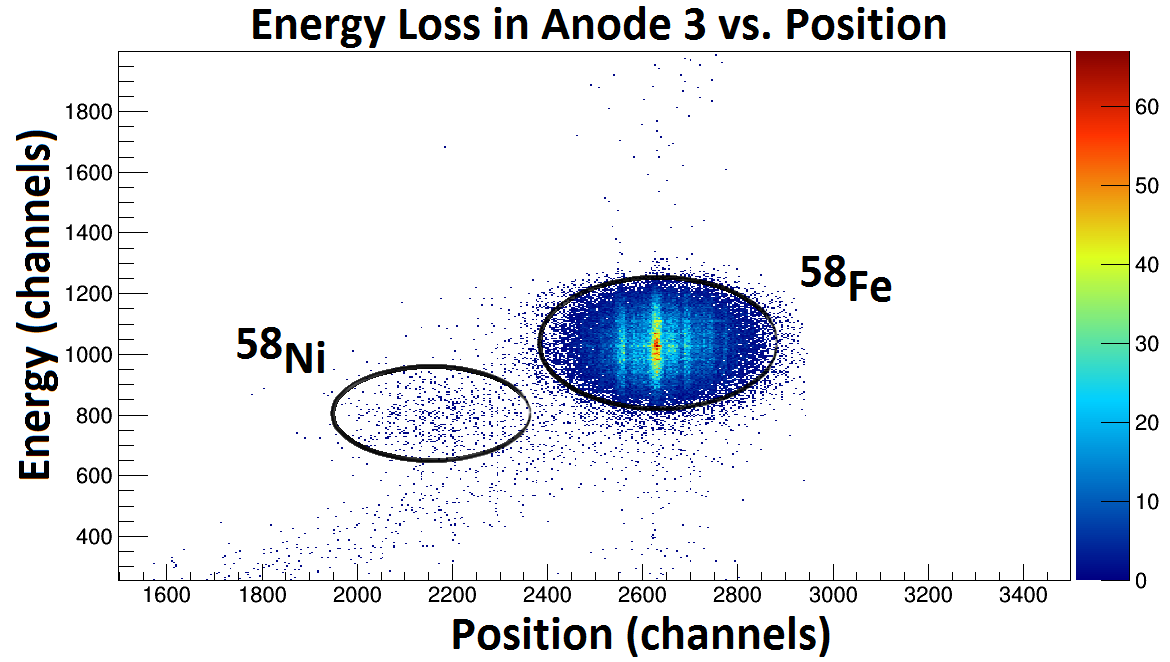}
            \caption{\small Blank dE3 vs. Position}    
            \label{fig:58de3}
        \end{subfigure}
        \quad
        \begin{subfigure}[b]{0.475\textwidth}   
            \centering 
            \includegraphics[width=\textwidth]{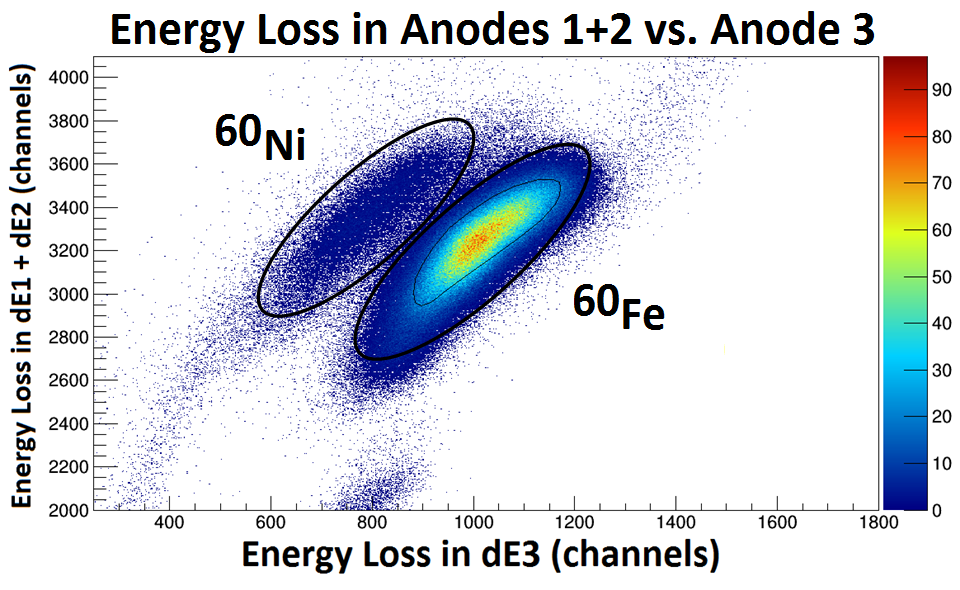}
            \caption{\small Fe-1 dE1 + dE2 vs. dE3, cuts drawn.}    
            \label{fig:Fe1drawn}
        \end{subfigure}
        \caption{\small The left column illustrates the crossover technique with Bragg Curve spectroscopy with a mass 58 beam, \isotope{Ni}{58} and \isotope{Fe}{58}. Mass 58 is a good approximation for the behavior observed at mass 60. The position of the beam particles as they exit the Spectograph magnet is recorded by a Parallel Grid Avalanche Counter (PGAC) detector and plotted on the x-axis in the left column plots. Following the PGAC is an ionization chamber (IC), split into 4 anodes. Each anode records the amount of energy deposited. In \ref{fig:58de1}, the energy deposited in the first anode is plotted on the y-axis. Here the isobar \isotope{Ni}{58} losses energy at a higher rate compared to \isotope{Fe}{58}. In \ref{fig:58de3}, the energy deposited in the third anode of the IC is plotted. Here \isotope{Fe}{58} is losing energy at a higher rate compared to \isotope{Ni}{58}. The right column shows the energy loss of anodes 1 and 2 plotted against the energy loss in anode 3. \ref{fig:Fe4drawn} is data for the Fe-4 material and \ref{fig:Fe1drawn} is data for the Fe-1 material. By using the crossover technique, good separation between the isobars of nickel and iron is observed.}
        \label{fig:oneAMS}
    \end{figure*}

For this work, the isotopic ratio of the Fe-4 material as published by Wallner et al. \cite{Wallner2015} was used as a reference value. This allows for a relative measurement of the isotopic ratio of the Fe-1 material. Examples of the AMS data taken in May 2016 are shown in Figure \ref{fig:oneAMS}. Several sets of measurements were made on each of the samples with periodic background measurements on material devoid of {\fesixty} (blanks). After determining the raw isotopic ratio of {\fesixty}/{\fe} of the reference, Fe-4, the absolute efficiency of beam transport was found and applied to the raw isotopic ratio of Fe-1. This is shown in Table \ref{table:Data}, giving an average {\fesixty}/{\fe} isotopic ratio for the Fe-1 sample of $(2.285 \pm 0.222) \times 10^{-6}$, shown in Table \ref{tab:Results}. Here, the uncertainty of the mean is calculated from the isotopic ratios of the three sets, respectively. Isotopic ratios are calculated with the following equation, where the Beam Trans. is the transport efficiency from the ion source Faraday cup to the beam line Faraday cup (accounting for the beam charge state): 
\begin{center}
\begin{eqnarray}
\text{\fesixty{}/\fe{}}=\left(\frac{\text{\fesixty{} Counts}}{\text{second}}\right )_{\text{detector}}\times \left (\frac{100}{\text{Beam Trans.\%}}\right )\times \left (\frac{\text{second}}{\text{\fe{} particles}}\right )_ {\text{ion source}} 
\end{eqnarray}
\end{center}

\clearpage

\clearpage
\begin{table}[h]
\centering
\begin{threeparttable}
\caption{Results of the Accelerator Mass Spectrometry measurement performed on the samples Fe-4 and Fe-1. Uncertainties indicated for the normalized {\fesixty}/{\fe} concentration for the Fe-1 sample are systematic uncertainties as calcuated from Table \ref{tab:Results}. }
\label{table:Data}
{\renewcommand{\arraystretch}{1.75}%
{\setlength{\tabcolsep}{1em}
\begin{tabular}{| c | c | c | c | c | c | c | c |}
\hline
Set                & Sample & \begin{tabular}[c]{@{}c@{}}Avg. {\fe}$^{-}$ \\ Current (nA)\end{tabular} & Time (s) & \begin{tabular}[c]{@{}c@{}}Trans.\\ {\fe}$^{-}$$\rightarrow$\\{\fe}$^{+16}$\\ (\%)\end{tabular} & \begin{tabular}[c]{@{}c@{}}Raw Counts \\ in Region\end{tabular} & \begin{tabular}[c]{@{}c@{}}Measured \\ {\fesixty}/{\fe}\\ ($\times 10^{-10})$ \end{tabular} & \begin{tabular}[c]{@{}c@{}}Normalized  \\ {\fesixty}/{\fe} \\($\times 10^{-9})$ \tnote{a}\end{tabular} \\ \hline
\multirow{3}{*}{1} & Blank  & 201.4                     & 1677.8   & 0.68             & 1                   &\textless0.0008 & \textless 0.001                                                       \\ \cline{2-7} 
                   & Fe-4   & 77.2                      & 2569.4   & 0.74              & 1156               &1.258 & $(2.082 \pm 0.091)$ \tnote{c}                                        \\ \cline{2-7} 
                   & Fe-1   & 101.1                     & 467.8    & 0.73               & 344633             &1599 & $(2645 \pm 145)$                                           \\ \hline
\multirow{2}{*}{2 \tnote{b}} & Fe-4   & 25.8                      & 1456.7   & 0.78               & 1232               &6.750 & $(2.082 \pm 0.091)$ \tnote{c}                                        \\ \cline{2-7} 
                   & Fe-1   & 57.5                      & 425.5    & 0.80               & 746016           & 6094  & $(1880 \pm 102)$                                          \\ \hline
\multirow{2}{*}{3 \tnote{b}} & Fe-4   & 75.7                      & 1417.3   & 0.75             & 3226            &  6.439  & $(2.082 \pm 0.091)$ \tnote{c}                                        \\ \cline{2-7} 
                   & Fe-1   & 62.8                      & 64.9     & 0.75               & 137621             &7205 & $(2330 \pm 115)$                                          \\ \hline

\end{tabular}}}
\begin{tablenotes}

\item[a] The background has been subtracted for the samples Fe-4 and Fe-1.
\item[b] Sets 2 and 3 have an increased beam transmission compared to Set 1.
\item[c] Reference value for Fe-4 from of Wallner et al. \cite{Wallner2015}, using $(1.145 \pm 0.050) \times 10^{12}$ {\fesixty} atoms in Fe-4 and $(4.95 \pm 0.01) \times 10^{20}$ \isotope{Fe}{56} atoms added to Fe-4.
\end{tablenotes}

\end{threeparttable}
\end{table}

\begin{table}[h]
\centering
\caption{Accelerator Mass Spectrometry Measurement Systematic Uncertainty Calculation.}
\label{tab:Results}
{\renewcommand{\arraystretch}{1.75}
\begin{tabular}{|l|c|c|c|}
\hline
                                       \multicolumn{1}{|c|}{Quantity \footnote{Bold quantity is the final total.}}                         & Amount                 & Relative Error (\%) \footnote{Other contributors to the uncertainty, such as statistics, are negligible comparatively.}    \\ \hline

                                       Wallner et al. Fe-4 Isotope Ratio                    & $(2.082 \pm 0.091) \times 10^{-9}$ {\fesixty}/{\fe} & 4.37                  \\ \hline 
                                       Faraday Cup Readings                                  & x                      & 1.0                   \\ \hline 
                                       Stable Fe added to Original Fe-1 Sample               & $(4.95 \pm 0.01) \times 10^{20}$ {\fe} atoms & 0.20                  \\ \hline
                                        \multicolumn{1}{|c|}{\textbf{Mean AMS Isotope Ratio of Fe-1} }       & \textbf{$\mathbf{(2.285 \pm 0.222) \times 10^{-6}}$}  $\mathbf{^{60}\textbf{Fe}/^{56}\textbf{Fe}}$ & \textbf{9.72} \footnote{Percent Error of the uncertainty in the mean of the 3 measurement sets as shown in Table \ref{table:Data}, specifically the standard deviation divided by the square root of 3.}                 \\ \hline 
\end{tabular}}
\end{table}

\subsection{Half-Life Comparison}

Knowing the isotopic ratio of {\fesixty}/{\fe} and the amount of {\fe} added to the sample, the total number of {\fesixty} atoms can be calculated. Relying on the dilution factors, Wallner et al. published a value of $(1.145 \pm 0.05)\times10^{15}$ {\fesixty} atoms in the full Fe-1 sample. In contrast, this work's direct measurement of the isotopic ratio of Fe-1 yields $(1.131 \pm 0.059)\times10^{15}$ {\fesixty} atoms in the full Fe-1 material, relying on Fe-4 as an AMS reference. Combining both of these numbers with this work's direct activity measurement from Section \ref{Activity} gives a half-life value of $(2.72 \pm 0.16)\times10^{6}$ years (for the Wallner isotopic ratio) and $(2.69 \pm 0.28)\times10^{6}$ years (for this work's isotopic ratio). Both results confirm the longer half-life value of Wallner et al. \cite{Wallner2015}. The first one indicates that the {\fesixty} activity measurement through the \isotope{Co}{60\text{m}} decay (this work) agrees with the one through the \isotope{Co}{60\text{g}} \cite{Wallner2015}. Although the second result also agrees with the longer half-life, it has a larger uncertainty due to the AMS measurement of this work.

\section{Results}

\begin{figure}[h]
\centering
\includegraphics[width=.75\textwidth]{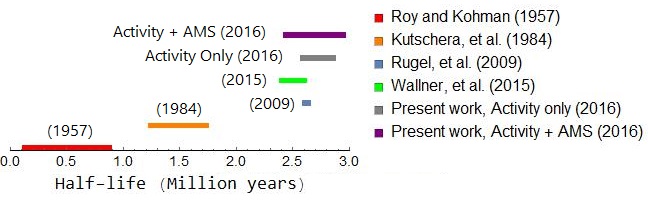}
\caption{All half-life measurements of {\fesixty} including this work. Note there is no y-axis. The individual measurements are separated out for the ease of the reader. The most recent measurements of Rugel et al. \cite{Rugel2009}, Wallner et al. \cite{Wallner2015}, and this work agree on a longer half-life than the previously accepted value of Kutschera et al. \cite{Kutschera1984}.}
\label{fig:Overlap}
\end{figure}

This work is the first to pair a direct decay measurement of \isotope{Co}{60\text{m}} with a corresponding AMS measurement. It is also the first to perform both on the same sample material, removing any reliance on a dilution or differing chemistry procedures. With the development of an {\fesixty} beam and a low-level-background counting station at the University of Notre Dame, we combined the results of the two experiments, finding a half-life of $(2.69 \pm 0.28) \times 10^{6}$ years. This is in agreement, albeit with a large uncertainty, with the most recent experiments (Rugel et al. \cite{Rugel2009} and Wallner et al. \cite{Wallner2015}) as illustrated in Figure \ref{fig:Overlap}. Combining this work's activity measurement with the AMS measurements performed instead by Wallner et al. gives a half-life value of $(2.72 \pm 0.16) \times 10^{6}$ years, also confirming a substantially longer half-life value than previously accepted. 

\section{Acknowledgements}
This work was funded by the National Science Foundation Grant Number PHY-1419765. We gratefully acknowledge Anton Wallner for his insight and help with this project, Max Bichler for the dilution series samples, Jiri Ulrich for performing the MCNP calculations, Vienna Environmental Research Accelerator Laboratory for making the samples available to us, and Nuclear Science Laboratory staff and technicians.

%
            
\end{document}